%% file: main.tex
\documentclass[conference]{IEEEtran}
\IEEEoverridecommandlockouts

\usepackage{cite}
\usepackage{amsmath,amssymb,amsfonts}
\usepackage{graphicx}
\usepackage{textcomp}
\usepackage{xcolor}
\usepackage{algorithm}
\usepackage{algpseudocode}
\usepackage{booktabs}
\usepackage{array}
\usepackage{multirow}
\usepackage[hidelinks]{hyperref}
\usepackage{orcidlink}
\usepackage{pbalance}
\usepackage{soul}

\begin{document}
\title{Centrality-based Structural-Electrical Contingency Screening and Transmission Reinforcement Prioritization \vspace{-0.25em}}
\bstctlcite{IEEEexample:BSTcontrol}

\author{%
\IEEEauthorblockN{%
Abanish~Tiwari\orcidlink{0009-0003-0609-8571},
Chandan~Chaudhary\orcidlink{0009-0002-2389-9568},~\emph{Student Member, IEEE},
Mekh~Raj~Joshi,\\
Mohammed~Ben-Idris\orcidlink{0000-0002-8731-8913},~\emph{Senior Member, IEEE},
and Joydeep~Mitra\orcidlink{0000-0001-9287-0983},~\emph{Fellow, IEEE}%
}
\IEEEauthorblockA{%
%Electrical and Computer Engineering, Michigan State %University, East Lansing, MI 48824, USA\\
E-mails: tiwariab@msu.edu;\; chaud152@msu.edu;\; mekh0095@gmail.com;\; benidris@msu.edu;\; mitraj@msu.edu%
}
\vspace{-2.4em}
}

\maketitle

\begin{abstract}
Rapid contingency screening is essential for secure transmission system operation and planning. Although full AC contingency analysis provides accurate system assessment, it requires high computational effort for large contingency sets. Topology-based screening methods reduce this burden but may not accurately represent operating-point-dependent electrical stress. In addition, conventional single-contingency screening does not quantify interactions among simultaneous outages or provide a direct criterion for reinforcement prioritization. This paper presents a hybrid structural--electrical screening approach that combines a computationally efficient, dispatch-independent structural analysis based on reactance-weighted centrality measures with a flow-based severity assessment using line outage distribution factors. Two indices are evaluated using the flow-based severity: a synergy index to quantify super-additive interactions among simultaneous line outages and a tolerance index to rank reinforcement candidates based on their reduction in aggregate post-contingency severity. The correspondence between structural indicators and electrical severity is evaluated quantitatively. Tests on the IEEE 39-bus system identify 115 super-additive line pairs beyond a 5\% threshold, with the most severe interaction exceeding seven times the linear combination of individual outage effects. The highest-ranked reinforcement reduces the mean post-contingency thermal severity by 7.5\%. The structural rankings show limited correlation with flow-based severity, with a Spearman coefficient of approximately 0.2. The proposed approach provides an efficient pre-screening method to reduce contingency and reinforcement search spaces before detailed AC and economic evaluations.

\end{abstract}


% =====================================================================
\section{Introduction}
% =====================================================================
\input{Contents/introduction}

% =====================================================================
\section{Methodology}
%\enlargethispage{2\baselineskip}
\label{sec:method}
% =====================================================================
\input{Contents/methodology}

% =====================================================================
%\vspace{-1em}
\section{Test System and Results}
\label{sec:results}
% =====================================================================
\input{Contents/results}

% =====================================================================
\section{Conclusion}
\label{sec:conclusion}
% =====================================================================
\input{Contents/conclusion}

\balance
% \enlargethispage{1\baselineskip}
\bibliographystyle{IEEEtran}
\bibliography{Contents/References}
\end{document}

%% file: Contents/introduction.tex
\IEEEPARstart{T}{he} secure operation of transmission networks requires continuous assessment of system vulnerability under component outages and varying operating conditions. The increasing size, interconnection level, and operational uncertainty of modern power systems have substantially expanded the number of credible contingency scenarios considered in both real-time operation and long-term planning. Conventional contingency analysis evaluates system response by solving nonlinear alternating-current (AC) power-flow equations for each outage condition and, when necessary, performing dynamic stability assessment. These methods provide accurate electrical information; however, their computational cost increases rapidly with the number and order of contingencies. Consequently, contingency screening techniques are widely employed to identify critical events using computationally efficient severity indicators before performing detailed analysis.

Traditional steady-state performance indices based on post-contingency line loading have been extensively used for contingency screening \cite{ejebe1979}, and subsequent studies have extended these approaches to real-time dynamic security assessment and remedial action evaluation \cite{mitra2017}. Transmission networks have also been represented as weighted graphs to identify critical components using concepts from complex-network theory. Centrality measures such as betweenness, closeness, and current-flow betweenness quantify path mediation, electrical proximity, and current redistribution, respectively \cite{freeman1977, freeman1978, NEWMAN200539}. With their low computational cost and independence from operating conditions, these measures have been applied to identify critical buses and transmission lines in large-scale power systems \cite{pagani2013, cuadra2015, coelho2019}. To improve the physical representation of graph-based models, researchers have incorporated electrical quantities such as line reactance and admittance into network formulations, leading to electrically weighted centrality measures and capability indicators \cite{wang2010, bompard2010}. Such methods enable efficient screening of large contingency sets without repeated power-flow solutions and are therefore well suited for planning studies \cite{albert2000}. In parallel, transmission expansion planning has been widely investigated using optimization-based and heuristic techniques to identify reinforcement strategies under technical and economic constraints \cite{latorre2003, hemmati2013}.

Despite these developments, several aspects of vulnerability assessment remain insufficiently explored. Structural metrics provide computationally efficient measures of network importance; nevertheless, their relationship with actual electrical severity depends on operating conditions, power transfers, and network constraints. Comparative studies have shown that topological importance and electrical criticality may differ significantly \cite{cotilla2012}, and further investigations have reported that topology-based models alone provide only limited representation of electricity-infrastructure vulnerability \cite{hines2010}. Although electrical weighting improves the physical interpretation of graph-based indices, the ability of these indices to represent operating-point-dependent contingency severity has not been comprehensively quantified. In addition, conventional screening primarily considers single-component outages, whereas major system disturbances often involve interacting or cascading outages. The loss of one transmission element redistributes power flow and may increase the severity of subsequent outages, producing compounded effects beyond the sum of individual contingencies \cite{motter2002}. Historical blackout analyses have likewise demonstrated the importance of cascading mechanisms and heavily stressed operating conditions in large-scale disturbances \cite{dobson2007}. Finally, although transmission expansion planning identifies effective reinforcement strategies, a quantitative measure that directly relates candidate line additions to reductions in outage vulnerability can provide valuable support for preliminary reinforcement screening.

This paper presents a structural--electrical approach that combines reactance-weighted centrality measures with line outage distribution factor (LODF)-based contingency severity analysis and quantitatively evaluates the correspondence between structural indicators and electrical severity. The main contributions of this work are summarized as follows:

\begin{itemize}
\item A hybrid structural--electrical approach is developed by combining
reactance-weighted centrality measures with LODF-based contingency severity
indices and evaluating their correspondence through statistical analysis.

\item A synergy index based on generalized line outage distribution factors is
introduced to quantify super-additive interactions among simultaneous outages and
identify higher-order contingency risks beyond N-1 assessment.

\item A tolerance index is formulated to rank transmission reinforcement
candidates according to their reduction in aggregate post-contingency severity,
providing an efficient screening criterion for expansion studies.
\end{itemize}

The proposed approach is intended to support detailed AC contingency analysis and transmission expansion planning by reducing the number of critical contingencies and reinforcement candidates requiring further evaluation. The remainder of this paper is organized as follows. Section~\ref{sec:method} presents the proposed methodology. Section~\ref{sec:results} discusses the numerical results obtained using the IEEE 39-bus test system. Section~\ref{sec:conclusion} concludes the paper and presents future research directions.

%% file: Contents/methodology.tex
The proposed approach evaluates contingencies and reinforcements using two
complementary analyses: a structural analysis based on reactance-weighted network
properties and a flow-based analysis based on operational severity. The latter is
used to quantify outage interactions and reinforcement benefits, while the
correspondence between the two analyses is evaluated statistically.

\subsection{System Model, Assumptions, and Notation}

The transmission system is represented as an undirected weighted graph
$G=(V,E,w)$, where $V$ denotes the set of $n$ buses, $E$ represents the set of
$m$ transmission lines, and $w$ defines the edge-weighting function. For a branch
$e$ connecting buses $u$ and $v$, the series reactance $x_e$ is used as the edge
weight for distance-based structural metrics, while the corresponding susceptance
$b_e=1/x_e$ is used in current-flow calculations. This weighting scheme enables the
graph representation to capture electrical proximity rather than only physical
connectivity.

The electrical severity assessment is performed using the DC power-flow model,
assuming a predominantly inductive network, voltage magnitudes close to unity, and
small phase-angle differences. The resulting flow-based indices are evaluated from
the base-case dispatch and branch parameters. A full AC power-flow solution based on
the Newton--Raphson method \cite{tinney1967} and implemented using MATPOWER
\cite{matpower2011} is used as an independent benchmark for validating the proposed
severity measures and contingency rankings. The proposed methodology requires only
branch reactances, thermal limits, and a base operating condition, while the
structural layer remains independent of dispatch conditions.

\subsection{Structural Layer}

The structural layer evaluates network vulnerability using three reactance-weighted
centrality measures: betweenness, closeness, and current-flow betweenness. These
metrics capture complementary structural characteristics of the transmission
network, including path dependency, electrical proximity, and current distribution.
Betweenness centrality identifies buses that participate in a large number of
minimum-reactance paths and is calculated using the dependency accumulation method
over weighted shortest paths \cite{freeman1977, brandes2001},

\begin{equation}
\hat{C}_B(j)=\frac{2}{(n-1)(n-2)}
\sum_{\substack{s,t\in V\setminus\{j\}\\s<t}}
\frac{\sigma_{st}(j)}{\sigma_{st}},
\label{eq:bc}
\end{equation}
where $\sigma_{st}$ denotes the number of minimum-reactance paths between buses
$s$ and $t$, and $\sigma_{st}(j)$ represents the number of paths passing through
bus $j$. The normalization provides a bounded metric suitable for comparison across
different systems. Closeness centrality measures the electrical proximity of a bus
to the remaining network using minimum-reactance distances,
\begin{equation}
C_C(j)=\frac{n_j-1}
{\sum_{t\in\mathcal{C}(j)\setminus\{j\}}d_{jt}},
\label{eq:cc}
\end{equation}
where $d_{jt}$ is the minimum accumulated reactance between buses $j$ and $t$ in
the connected component $\mathcal{C}(j)$ containing bus $j$.

Current-flow betweenness provides a closer representation of electrical behavior by
considering current redistribution through all available paths according to their
reactances \cite{NEWMAN200539, klein1993}. Using the weighted Laplacian matrix as follows,
\begin{equation}
L=A^{\top}\mathrm{diag}(b)A,
\end{equation}
where $A$ is the branch--bus incidence matrix, the voltage response caused by a unit
current injection between buses $s$ and $t$ is obtained using \eqref{voltresp} \cite{brandesfleischer2005}.
\begin{equation}\label{voltresp}
L\mathbf{v}^{(st)}=\mathbf{i}_{st}
\end{equation}
The current throughput associated with bus $j$ is defined as follows,
\begin{equation}
I_{st}(j)=\frac{1}{2}\sum_{r\sim j}
b_{jr}|v_j^{(st)}-v_r^{(st)}|,
\end{equation}
and the current-flow betweenness is calculated using \eqref{eq:cfbc}.
\begin{equation}
C_{CFB}(j)=\frac{2}{n_j(n_j-1)}
\sum_{s<t}I_{st}(j).
\label{eq:cfbc}
\end{equation}
The three measures are considered jointly because they represent different
structural characteristics. Betweenness identifies the role of a bus in minimum-reactance paths, closeness measures its electrical proximity to other buses, and current-flow betweenness characterizes current redistribution through multiple paths..

Following each contingency, connectivity is evaluated using the algebraic
connectivity $\lambda_2$, defined as the second-smallest eigenvalue of the graph
Laplacian. A positive $\lambda_2$ indicates that the network remains connected
\cite{fiedler1973}. Islanding contingencies are classified separately and excluded
from the valid contingency set $E_v$ to maintain well-defined structural distance
and centrality calculations. For a contingency $e$, the change in centrality at bus
$j$ is expressed as
\begin{equation}
\Delta C_i(j,e)=|C_i^{(e)}(j)-C_i(j)|,
\end{equation}
where $i\in\{B,C,CFB\}$. The aggregated structural impacts for lines and buses are
defined as follows,
\begin{equation}
\mathrm{GI}_i^{\mathrm{line}}(e)=\sum_{j\in V}\Delta C_i(j,e),
\end{equation}
\begin{equation}
\mathrm{GI}_i^{\mathrm{bus}}(j)=\sum_{e\in E_v}\Delta C_i(j,e).
\end{equation}

The line-wise impact provides a structural ranking of contingencies and is used as
a low-cost pre-screening metric, whereas the bus-wise impact identifies locations
whose structural importance varies significantly across the contingency set.

\subsection{Flow-Based Severity Layer}
The flow-based layer quantifies operational contingency severity using the base-case
dispatch and LODFs. LODFs represent the fraction
of the pre-contingency flow of an outaged line transferred to remaining lines and
enable efficient post-contingency analysis without repeated power-flow solutions
\cite{wood2014}. For the outage of line $e$, the post-contingency flow on line $l$
is obtained as follows,
\begin{equation}
f_l^{(e)}=f_l+\mathrm{LODF}_{l,e}f_e,
\label{postcon}
\end{equation}
where $f_l$ is the base-case flow and $\mathrm{LODF}_{l,e}$ is the corresponding
LODF. The thermal performance index is defined using \eqref{eq:pi}.
\begin{equation}
\mathrm{PI}_{\mathrm{th}}(e)=
\sum_{l\neq e}
\left(\frac{f_l^{(e)}}{f_l^{\max}}\right)^2 ,
\label{eq:pi}
\end{equation}
where $f_l^{\max}$ denotes the thermal limit of line $l$. The quadratic form,
adopted from conventional contingency performance indices
\cite{ejebe1979}, increases sensitivity to heavily loaded branches and provides a
measure of thermal stress severity.

For simultaneous outages, generalized LODFs are used to determine the resulting
post-contingency flows, provided that the removed lines do not cause network
islanding \cite{guler2007}. The corresponding thermal severity is calculated as
\begin{equation}
\mathrm{PI}_{\mathrm{th}}^{(e_1,e_2)}
=
\sum_{l\neq e_1,e_2}
\left(
\frac{f_l^{(e_1,e_2)}}{f_l^{\max}}
\right)^2 .
\end{equation}

Since the DC formulation does not include reactive-power effects and voltage
magnitude limitations, an AC-based voltage
severity index is included as an independent benchmark. The maximum voltage
stability index obtained from the AC solution is defined as follows \cite{kessel1986},
\begin{equation}
\mathrm{PI}_{v}(e)=\max_{j\in V}L_j^{(e)},
\end{equation}
where $L_j^{(e)}$ represents the voltage stability index of bus $j$ following outage
$e$. This metric evaluates the ability of the thermal severity measure to identify
voltage-related vulnerabilities. Because flow-based severity depends on the
operating condition, the analysis is repeated over multiple operating points
generated through non-uniform load perturbations to evaluate ranking robustness
under varying power-flow patterns.
\begin{algorithm}[t]
\caption{Hybrid structural--electrical contingency and reinforcement screening}
\label{alg:pipeline}
\begin{algorithmic}[1]

\State \textbf{Input:} Weighted graph $G$, branch ratings, and base-case dispatch

\State Compute structural centralities $C_B$, $C_C$, and $C_{CFB}$

\For{each single-line outage $e$}

    \State Evaluate connectivity using algebraic connectivity $\lambda_2$

    \If{$\lambda_2>0$}

        \State Compute structural impact $\mathrm{GI}^{\mathrm{line}}(e)$
        and add $e$ to valid outage set $E_v$

        \State Evaluate flow-based severities
        $\mathrm{PI}_{\mathrm{th}}(e)$ and $\mathrm{PI}_{v}(e)$

    \Else

        \State Classify $e$ as an islanding contingency

    \EndIf

\EndFor
\State Quantify structural--electrical correlation using Spearman rank coefficient

\For{each valid outage pair $(e_1,e_2)$}

    \State Compute synergy index using \eqref{eq:synergy}
\EndFor
\For{each reinforcement candidate $c$}

    \State Compute tolerance index using \eqref{eq:tolerance}
\EndFor
\State \textbf{Output:} Ranked contingencies, critical outage pairs, and reinforcement candidates
\end{algorithmic}
\end{algorithm}

\subsection{Synergy Index for Compounded Severity}

Single-contingency severity analysis identifies critical individual outages but does not quantify interactions among simultaneous outages. To capture such effects, the
synergy index is defined on an additive severity scale \cite{knol2011} by comparing
the joint contingency severity increment with the sum of individual increments,
\begin{equation}
\mathrm{Synergy}(e_1,e_2)=
\frac{\Delta\mathrm{PI}_{\mathrm{th}}^{(e_1,e_2)}}
{\Delta\mathrm{PI}_{\mathrm{th}}(e_1)+
\Delta\mathrm{PI}_{\mathrm{th}}(e_2)},
\label{eq:synergy}
\end{equation}
where $\Delta\mathrm{PI}_{\mathrm{th}}(\cdot)$ represents the increase in thermal
severity relative to the pre-contingency condition. The index is dimensionless,
with values greater than unity indicating super-additive interactions, where the
combined outage produces a severity increase larger than the linear contribution
of individual outages.

The significance of the synergy index is that it identifies hidden higher-order contingency risks that cannot be revealed through conventional N-1 screening. Although individual lines may exhibit low severity when evaluated independently, their simultaneous outage can create critical conditions due to shared transfer corridors or sequential flow redistribution effects. Therefore, the index captures the structural--electrical interaction between outages rather than treating their effects as independent contributions. In practice, it reduces the large space of multi-line contingencies to a ranked subset of super-additive combinations suitable
for detailed analysis, protection studies, or contingency planning. The formulation can be extended to higher-order contingencies by replacing the numerator and denominator with the corresponding $k$-line severity terms; however, since the number of combinations increases as $\binom{m}{k}$, the structural layer is used to pre-screen candidate combinations before evaluating the flow-based severity.
Because most line pairs are electrically weakly coupled, the synergy index generally approaches unity; therefore, a pair is classified as super-additive only when its
value exceeds unity by a predefined tolerance margin $\tau$.

\subsection{Tolerance Index for Reinforcement}

The impact of candidate reinforcements is quantified through their ability to reduce aggregate post-contingency severity. Let $\overline{\mathrm{PI}}_{\mathrm{th}}(G)$ denote the average thermal severity over all valid single-line contingencies of network $G$. For a candidate line $c$ added between existing buses, forming the reinforced network $G_c$, the tolerance index is
defined as
\begin{equation}
\mathrm{Tolerance}(c)=
\frac{\overline{\mathrm{PI}}_{\mathrm{th}}(G)-
\overline{\mathrm{PI}}_{\mathrm{th}}(G_c)}
{\overline{\mathrm{PI}}_{\mathrm{th}}(G)}
\times 100\%.
\label{eq:tolerance}
\end{equation}

The contingency set is kept identical for the original and reinforced networks, and the added line is not considered as an outage, ensuring that the severity difference is attributed only to the reinforcement effect. A positive tolerance index indicates that the candidate improves system robustness by reducing post-contingency stress. Unlike conventional expansion criteria focused primarily on capacity adequacy, this metric directly evaluates vulnerability reduction and provides an interpretable measure for ranking candidate reinforcements. High-ranking candidates represent branches that provide effective alternative transfer paths and mitigate severe contingency-induced overloads. Since complete transmission expansion evaluation requires consideration of cost, feasibility, right-of-way, voltage constraints, and stability requirements, the proposed index is intended as a computationally efficient pre-screening tool that reduces the candidate space before detailed planning studies. Together with the synergy index, it completes the analysis by identifying critical outage interactions and ranking reinforcements that provide the greatest reduction in network vulnerability.

\subsection{Coupling and Validation}

The relationship between structural vulnerability and electrical severity is quantified by explicitly coupling the structural and flow-based layers \cite{hines2010}. For valid contingencies, the agreement between structural rankings and flow-based severity rankings is evaluated using the Spearman rank-correlation coefficient \cite{spearman1904}, together with bootstrap confidence intervals and permutation tests against the null hypothesis of independent rankings. The same correlation analysis is applied to evaluate the relationship between structural and flow-based synergy and tolerance measures.

The proposed indices are further validated as quantitative measures. The closed-form synergy index is compared with exact
DC re-solutions of simultaneous outages and with AC power-flow solutions, while the tolerance index is compared against the change in aggregate contingency severity obtained from full AC re-solutions of reinforced networks. The validation performance is quantified using the maximum deviation from the exact solution and the correlation with the AC benchmark. The reference calculations, including the Newton-Raphson AC power flow and direct-current contingency solutions, are standard network analysis procedures \cite{tinney1967, wood2014, ejebe1979}; therefore, the proposed indices are validated against independent computational models.

In addition, the flow-based severity ranking is compared with a conventional alternating-current contingency ranking using rank correlation and critical-event recall. This validation establishes the operating range in which the structural layer can reliably reduce the search space and identifies cases where the flow-based layer must determine the final ranking. Thus, the proposed method replaces the assumption of topological adequacy with a quantitative assessment of structural--electrical consistency.

% \subsection{Computational Complexity}

% For a network with $n$ buses and $m$ branches, the structural layer evaluates betweenness and closeness centralities with computational complexity $O(nm+n^2\log n)$ for sparse weighted graphs \cite{brandes2001}. Current-flow betweenness requires a Laplacian pseudo-inversion with complexity $O(n^3)$, followed by bus throughput calculations with complexity $O(mn)$ \cite{brandesfleischer2005}. 

% In the flow-based layer, a single-line outage is represented as a rank-one update and a double-line outage as a rank-two update of the factorized DC power-flow system, requiring approximately $O(n^2)$ operations \cite{golub2013, guler2007}. This low-rank formulation enables efficient evaluation of large contingency sets because the computational cost of additional outages grows quadratically. Consequently, exhaustive second-order contingency screening with $\binom{m}{2}$ line pairs and reinforcement evaluation remain computationally tractable, while detailed flow-based analysis is performed only for candidates identified through the structural pre-screening stage.

%% file: Contents/results.tex
The proposed framework was evaluated on the IEEE 39-bus New England test system, which consists of 39 buses and 46 transmission branches. The MATPOWER case data were used for branch reactances, thermal limits, and the base-case operating condition. Line outage distribution factors, direct-current contingency solutions, and alternating-current power-flow solutions were obtained using MATPOWER, while the structural metrics were calculated using sparse graph-based formulations. The algebraic-connectivity screening identified 35 connected single-line contingencies and 11 islanding cases. For double-line contingencies, 562 of the $\binom{46}{2}=1035$ possible outage pairs remained electrically connected and were considered for severity evaluation. The complete analysis, including eight operating points and 75 reinforcement candidates, was completed in less than one minute on a standard eight-core workstation.

\begin{figure}[t]
\centering
\includegraphics[width=\columnwidth]{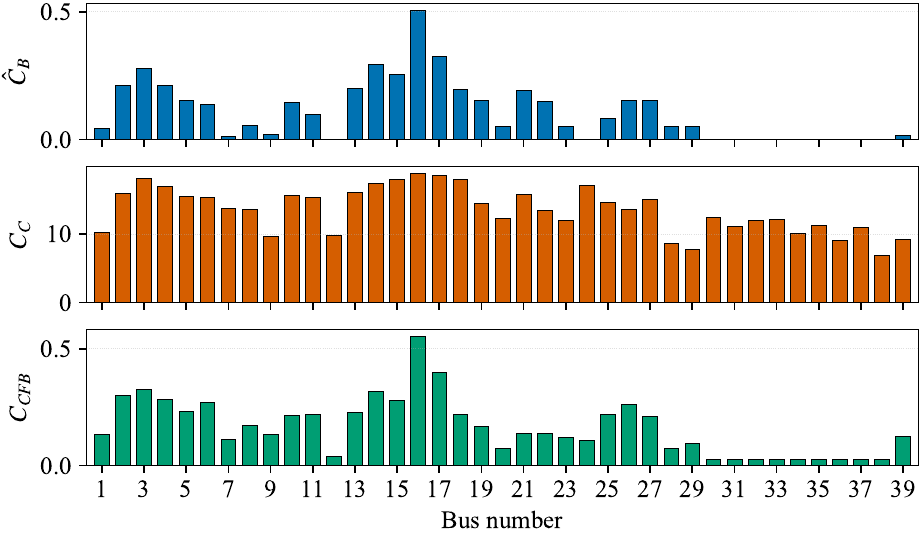}
\vspace{-2em}
\caption{Normalized bus centrality profiles for the reactance-weighted structural
measures.}
\label{fig:base}
\end{figure}

\begin{figure}[t]
\centering
\includegraphics[width=1\columnwidth]{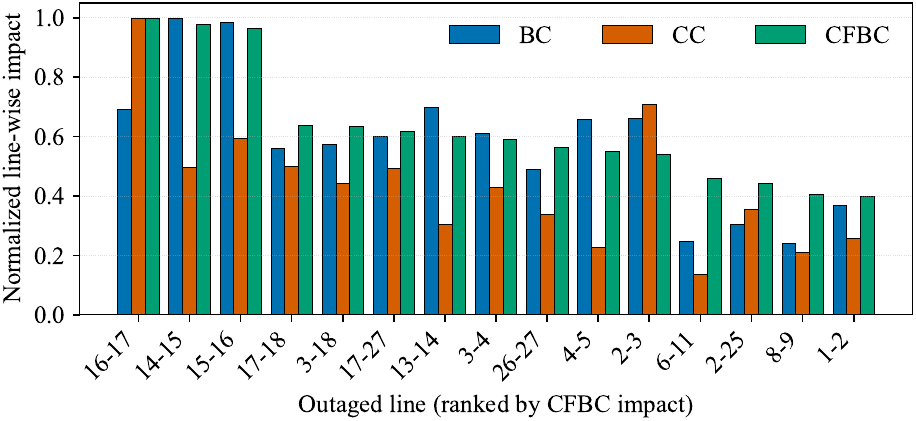}
\vspace{-2em}
\caption{Structural impact ranking of the most disruptive single-line outages.}
\label{fig:line}
\end{figure}

\begin{figure}[thb]
\centering
\includegraphics[width=1\columnwidth]{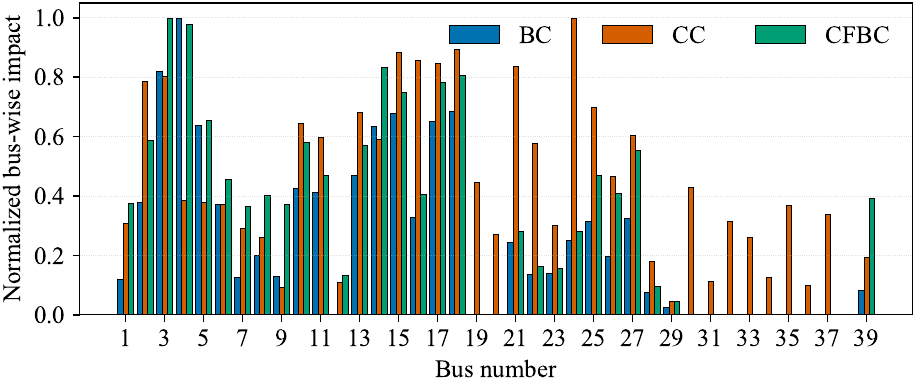}
\vspace{-2em}
\caption{Normalized bus structural impact across the contingency set.}
\label{fig:bus}
\end{figure}

The structural analysis was first evaluated using the three reactance-weighted centrality measures. Fig.~\ref{fig:base} shows the normalized centrality profiles for all buses. The three measures consistently identify buses 16 and 17 and the corridor between buses 14--17 as structurally dominant regions. Similarly, Fig.~\ref{fig:line} shows that outages producing the largest structural redistribution are concentrated in these highly connected transmission corridors. The bus-level structural impact shown in Fig.~\ref{fig:bus} further confirms that the same locations experience the largest changes in network centrality after contingencies.

\begin{figure}[tb]
\centering
\includegraphics[width=0.8\columnwidth]{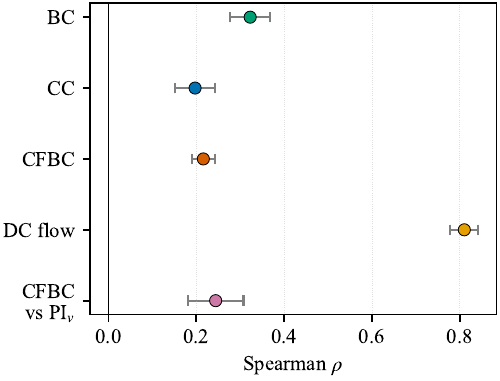}
\vspace{-1em}
\caption{Spearman correlation between structural rankings and electrical severity
across operating conditions.}
\label{fig:coupling}
\end{figure}

The relationship between the structural rankings and electrical contingency severity is evaluated in Fig.~\ref{fig:coupling} using the Spearman rank correlation coefficient over eight operating conditions. Each operating point was generated by independently perturbing the base-case load at each bus within $\pm30\%$ and proportionally rescaling generation to maintain active-power balance, thereby producing different power-flow distributions and
contingency severity rankings. The result illustrates that the current-flow betweenness exhibits a correlation of $0.22\pm0.03$, whereas betweenness and closeness centralities achieve correlations of $0.32$ and $0.20$, respectively. By comparison, the ranking derived from DC line flows, included as a reference, attains a substantially higher correlation of $0.81$, while the correlation with the AC voltage-stability index remains comparatively low at $0.24$. 

% These results indicate that reactance-weighted structural metrics have limited capability to represent operating-point-dependent contingency severity, since electrical loading is governed by the distribution of power injections, network constraints, and power-flow redistribution rather than network topology alone. This observation is consistent with previous studies reporting the limited fidelity of topology-based vulnerability assessment for power systems \cite{hines2010}. Accordingly, the structural metrics are employed for computationally efficient preliminary screening, whereas the flow-based thermal severity serves as the basis for contingency ranking and for evaluating the proposed synergy and tolerance indices.

\begin{figure}[h]
\centering
\includegraphics[width=0.8\columnwidth]{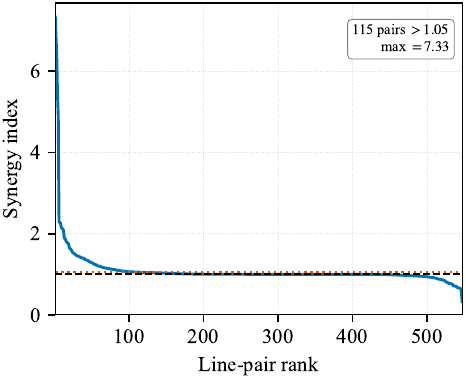}
\vspace{-1em}
\caption{Ranked distribution of the flow-based synergy index for valid line pairs.}
\label{fig:synergy}
\end{figure}

The compounded severity of simultaneous outages was evaluated using the synergy
index in \eqref{eq:synergy}. Fig.~\ref{fig:synergy} shows the ranked distribution
of synergy values for all valid line pairs. Most outage pairs exhibit values close to unity, indicating nearly additive behavior. However, a limited number of pairs show significant super-additive interactions. Using a 5\% threshold above unity, 115 out of 547 evaluated pairs were identified as super-additive, with 26 pairs exceeding a synergy value of 1.5. The maximum observed synergy was 7.33, produced by the simultaneous outage of lines 5--6 and 6--7. These results demonstrate that certain combinations of individually low-severity outages can produce severe system stress when considered jointly.

\begin{table}[!t]
\centering
\caption{Highest flow-based synergy-index line pairs.}
\label{tab:synergy}
\begin{tabular}{@{}cccc@{}}
\toprule
\textbf{Line 1} & \textbf{Line 2} & \textbf{Synergy} & \textbf{Rank}\\
\midrule
5--6  & 6--7   & 7.33 & 1 \\
8--9  & 14--15 & 6.82 & 2 \\
5--6  & 7--8   & 6.35 & 3 \\
5--8  & 6--7   & 5.43 & 4 \\
5--8  & 7--8   & 4.89 & 5 \\
1--2  & 2--3   & 2.30 & 6 \\
4--14 & 6--11  & 2.27 & 7 \\
15--16& 17--18 & 2.27 & 8 \\
4--14 & 16--17 & 2.16 & 9 \\
4--5  & 13--14 & 2.13 & 10 \\
\bottomrule
\end{tabular}
\end{table}

The reinforcement screening was performed using the tolerance index defined in \eqref{eq:tolerance}. Candidate lines were generated between two-hop neighboring buses, with median existing reactance and thermal rating assigned to each candidate. Table~\ref{tab:reinf} lists the ten highest-ranked reinforcement options. The addition of line 4--13 produced the largest reduction in average contingency
severity, with a tolerance value of 7.49\%, followed by lines 6--10 and 3--25. Among the 75 evaluated candidates, 65 reduced the aggregate contingency severity, while the improvement was concentrated among the highest-ranked reinforcements.

\begin{table}[tbh]
\centering
\caption{Highest flow-based tolerance reinforcement candidates.}
\label{tab:reinf}
\begin{tabular}{@{}clc@{}}
\toprule
\textbf{Rank} & \textbf{Candidate line} & \textbf{Tolerance (\%)}\\
\midrule
1  & 4--13  & 7.49 \\
2  & 6--10  & 6.91 \\
3  & 3--25  & 6.53 \\
4  & 7--11  & 6.02 \\
5  & 16--33 & 5.81 \\
6  & 10--14 & 5.77 \\
7  & 15--19 & 5.66 \\
8  & 5--11  & 5.03 \\
9  & 17--19 & 4.99 \\
10 & 16--22 & 4.87 \\
\bottomrule
\end{tabular}
\end{table}

\begin{figure}[h]
\centering
\includegraphics[width=0.8\columnwidth]{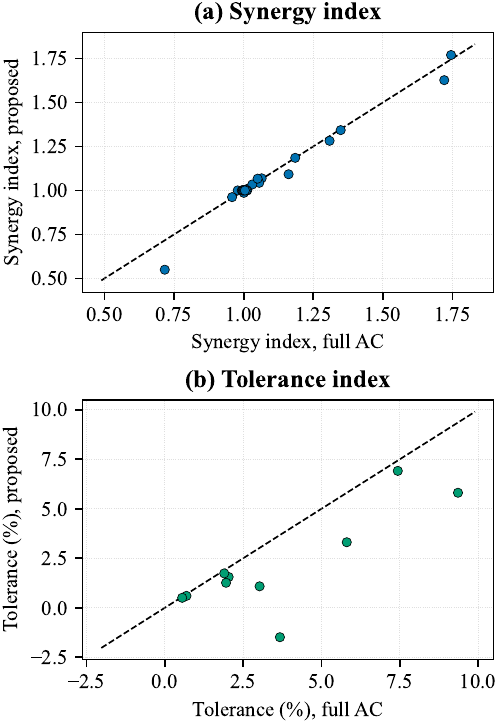}
\vspace{-1em}
\caption{Validation of synergy and tolerance indices against AC power-flow.}
\label{fig:validation}
\end{figure}

The proposed indices are validated against independent DC and AC network solutions to assess their numerical accuracy and ranking consistency. The synergy index, computed using generalized LODFs, produces results similar to those obtained from the DC re-solution, with a  deviation of nearly zero and achieves a Pearson correlation coefficient of $0.998$ with full AC re-solutions. Similarly, the tolerance index accurately estimates the reduction in aggregate post-contingency thermal severity, yielding a Pearson correlation coefficient of $0.95$ and a mean absolute deviation of approximately $1\%$. The parity plots in Fig.~\ref{fig:validation} confirm the close agreement between the proposed indices and the AC reference solutions.

The proposed methodology is further benchmarked against conventional AC contingency analysis. The LODF-based thermal severity ranking achieves a Spearman rank correlation coefficient of $0.99$ with the full AC contingency ranking and correctly identifies all of the most critical outages. For higher-order contingencies, the synergy index attains a Spearman rank correlation coefficient of $0.83$ with full AC $N$--2 analysis and identifies $83\%$ of the highest-ranked line pairs. Likewise, the tolerance index achieves a Spearman rank correlation coefficient of $0.81$ with the AC reinforcement ranking and correctly identifies all of the highest-ranked reinforcement candidates. These results demonstrate that the indices provide computationally efficient and accurate surrogates for contingency severity and reinforcement assessment. The results demonstrate that the proposed synergy and tolerance indices accurately quantify higher-order contingency interactions and reinforcement benefits, respectively, providing practical metrics for vulnerability assessment and candidate prioritization.

%% file: Contents/conclusion.tex
This paper has presented a hybrid structural-electrical method for transmission contingency screening and reinforcement assessment. The method integrated a dispatch-independent structural layer based on reactance-weighted graph metrics with an electrical severity layer based on line outage distribution factors. The relationship between structural vulnerability and operational severity was quantified without assuming direct equivalence between the two. The results showed that structural metrics provided an efficient mechanism for identifying vulnerable network regions and reducing the contingency search space. However, electrical assessment remained necessary to capture operating-condition-dependent severity and power-flow redistribution effects. The proposed synergy and tolerance indices extended conventional contingency analysis by identifying compounded outage interactions and ranking reinforcement candidates based on their impact on system security. The developed method provided a scalable screening approach that combined computational efficiency with electrical relevance and could support detailed AC, dynamic, and economic security studies. Future work will focus on evaluating larger transmission systems, incorporating injection-aware structural measures, including AC and dynamic security constraints, and integrating the proposed indices with reliability-based and economic transmission expansion planning extended to a larger and more complex system.

%% file: Contents/References.bib
@IEEEtranBSTCTL{IEEEexample:BSTcontrol,
  CTLdash_repeated_names = "no"
}

@article{coelho2019,
  author  = {E. P. R. Coelho and M. H. M. Paiva and M. E. V. Segatto and G. Caporossi},
  title   = {A New Approach for Contingency Analysis Based on Centrality Measures},
  journal = {IEEE Systems Journal},
  volume  = {13},
  number  = {2},
  pages   = {1915--1923},
  year    = {2019},
  month   = {jun},
  doi     = {10.1109/JSYST.2018.2881558}
}

@article{knol2011,
  author  = {M. J. Knol and T. J. VanderWeele and R. H. Groenwold and O. H. Klungel and M. M. Rovers and D. E. Grobbee},
  title   = {Estimating measures of interaction on an additive scale for preventive exposures},
  journal = {Eur. J. Epidemiol.},
  volume  = {26},
  number  = {6},
  pages   = {433--438},
  year    = {2011},
  month   = {jun},
  doi     = {10.1007/s10654-011-9554-9}
}

@article{freeman1977,
  author  = {Freeman, Linton C.},
  title   = {A Set of Measures of Centrality Based on Betweenness},
  journal = {Sociometry},
  volume  = {40},
  number  = {1},
  pages   = {35--41},
  year    = {1977},
  doi     = {10.2307/3033543}
}

@article{freeman1978,
  author  = {Freeman, Linton C.},
  title   = {Centrality in social networks conceptual clarification},
  journal = {Social Networks},
  volume  = {1},
  number  = {3},
  pages   = {215--239},
  year    = {1978},
  doi     = {10.1016/0378-8733(78)90021-7}
}

@article{NEWMAN200539,
title = {A measure of betweenness centrality based on random walks},
journal = {Social Networks},
volume = {27},
number = {1},
pages = {39-54},
year = {2005},
issn = {0378-8733},
doi = {10.1016/j.socnet.2004.11.009},
author = {M.E. J. Newman},
}

@article{brandes2001,
  author  = {U. Brandes},
  title   = {A Faster Algorithm for Betweenness Centrality},
  journal = {Journal of Mathematical Sociology},
  volume  = {25},
  number  = {2},
  pages   = {163--177},
  year    = {2001},
  doi     = {10.1080/0022250X.2001.9990249}
}

@inproceedings{brandesfleischer2005,
  author    = {U. Brandes and D. Fleischer},
  title     = {Centrality Measures Based on Current Flow},
  booktitle = {Proc. 22nd Annu. Symp. Theoretical Aspects of Computer Science (STACS), LNCS 3404},
  pages     = {533--544},
  year      = {2005},
  doi       = {10.1007/978-3-540-31856-9_44}
}

@inproceedings{wang2010,
  author    = {Z. Wang and A. Scaglione and R. J. Thomas},
  title     = {Electrical Centrality Measures for Electric Power Grid Vulnerability Analysis},
  booktitle = {49th IEEE Conf. Decision and Control (CDC)},
  pages     = {5792--5797},
  year      = {2010},
  doi       = {10.1109/CDC.2010.5717964}
}

@article{bompard2010,
  author  = {E. Bompard and R. Napoli and F. Xue},
  title   = {Extended Topological Approach for the Assessment of Structural Vulnerability in Transmission Networks},
  journal = {IET Generation, Transmission \& Distribution},
  volume  = {4},
  number  = {6},
  pages   = {716--724},
  year    = {2010},
  doi     = {10.1049/iet-gtd.2009.0452}
}

@article{pagani2013,
  author  = {G. A. Pagani and M. Aiello},
  title   = {The Power Grid as a Complex Network: A Survey},
  journal = {Physica A: Statistical Mechanics and its Applications},
  volume  = {392},
  number  = {11},
  pages   = {2688--2700},
  year    = {2013},
  doi     = {10.1016/j.physa.2013.01.023}
}

@article{guler2007,
  author  = {T. G\"{u}ler and G. Gross and M. Liu},
  title   = {Generalized Line Outage Distribution Factors},
  journal = {IEEE Transactions on Power Systems},
  volume  = {22},
  number  = {2},
  pages   = {879--881},
  year    = {2007},
  month   = {may},
  doi     = {10.1109/TPWRS.2006.888950}
}

@article{ejebe1979,
  author  = {G. C. Ejebe and B. F. Wollenberg},
  title   = {Automatic Contingency Selection},
  journal = {IEEE Transactions on Power Apparatus and Systems},
  volume  = {PAS-98},
  number  = {1},
  pages   = {97--109},
  year    = {1979},
  month   = {jan},
  doi     = {10.1109/TPAS.1979.319518}
}

@article{spearman1904,
  author  = {C. Spearman},
  title   = {The Proof and Measurement of Association between Two Things},
  journal = {The American Journal of Psychology},
  volume  = {15},
  number  = {1},
  pages   = {72--101},
  year    = {1904},
  doi     = {10.2307/1412159}
}

@book{wood2014,
  author    = {A. J. Wood and B. F. Wollenberg and G. B. Shebl\'{e}},
  title     = {Power Generation, Operation, and Control},
  edition   = {3rd},
  publisher = {Wiley},
  address   = {Hoboken, NJ, USA},
  year      = {2014}
}

@article{klein1993,
  author  = {D. J. Klein and M. Randi\'{c}},
  title   = {Resistance Distance},
  journal = {Journal of Mathematical Chemistry},
  volume  = {12},
  pages   = {81--95},
  year    = {1993},
  doi     = {10.1007/BF01164627}
}

@article{cotilla2012,
  author  = {E. Cotilla-Sanchez and P. D. H. Hines and C. Barrows and S. Blumsack},
  title   = {Comparing the Topological and Electrical Structure of the North American Electric Power Infrastructure},
  journal = {IEEE Systems Journal},
  volume  = {6},
  number  = {4},
  pages   = {616--626},
  year    = {2012},
  doi     = {10.1109/JSYST.2012.2183033}
}

@article{fiedler1973,
  author  = {M. Fiedler},
  title   = {Algebraic Connectivity of Graphs},
  journal = {Czechoslovak Mathematical Journal},
  volume  = {23},
  number  = {2},
  pages   = {298--305},
  year    = {1973},
  doi     = {10.21136/CMJ.1973.101168}
}

@article{albert2000,
  author  = {R. Albert and H. Jeong and A.-L. Barab\'{a}si},
  title   = {Error and Attack Tolerance of Complex Networks},
  journal = {Nature},
  volume  = {406},
  number  = {6794},
  pages   = {378--382},
  year    = {2000},
  doi     = {10.1038/35019019}
}

@article{latorre2003,
  author  = {G. Latorre and R. D. Cruz and J. M. Areiza and A. Villegas},
  title   = {Classification of Publications and Models on Transmission Expansion Planning},
  journal = {IEEE Transactions on Power Systems},
  volume  = {18},
  number  = {2},
  pages   = {938--946},
  year    = {2003},
  doi     = {10.1109/TPWRS.2003.811168}
}

@article{kessel1986,
  author  = {P. Kessel and H. Glavitsch},
  title   = {Estimating the Voltage Stability of a Power System},
  journal = {IEEE Transactions on Power Delivery},
  volume  = {1},
  number  = {3},
  pages   = {346--354},
  year    = {1986},
  month   = {jul},
  doi     = {10.1109/TPWRD.1986.4308013}
}

@article{hines2010,
  author  = {P. Hines and E. Cotilla-Sanchez and S. Blumsack},
  title   = {Do Topological Models Provide Good Information about Electricity Infrastructure Vulnerability?},
  journal = {Chaos: An Interdisciplinary Journal of Nonlinear Science},
  volume  = {20},
  number  = {3},
  pages   = {033122},
  year    = {2010},
  doi     = {10.1063/1.3489887}
}

@article{cuadra2015,
  author  = {L. Cuadra and S. Salcedo-Sanz and J. Del Ser and S. Jim\'{e}nez-Fern\'{a}ndez and Z. W. Geem},
  title   = {A Critical Review of Robustness in Power Grids Using Complex Networks Concepts},
  journal = {Energies},
  volume  = {8},
  number  = {9},
  pages   = {9211--9265},
  year    = {2015},
  doi     = {10.3390/en8099211}
}

@article{mitra2017,
  author  = {J. Mitra and M. Benidris and N. Nguyen and S. Deb},
  title   = {A Visualization Tool for Real-Time Dynamic Contingency Screening and Remedial Actions},
  journal = {IEEE Transactions on Industry Applications},
  volume  = {53},
  number  = {4},
  pages   = {3268--3278},
  year    = {2017},
  month   = {jul},
  doi     = {10.1109/TIA.2017.2686353}
}

@article{motter2002,
  author  = {A. E. Motter and Y.-C. Lai},
  title   = {Cascade-Based Attacks on Complex Networks},
  journal = {Physical Review E},
  volume  = {66},
  number  = {6},
  pages   = {065102},
  year    = {2002},
  doi     = {10.1103/PhysRevE.66.065102}
}

@article{dobson2007,
  author  = {I. Dobson and B. A. Carreras and V. E. Lynch and D. E. Newman},
  title   = {Complex Systems Analysis of Series of Blackouts: Cascading Failure, Critical Points, and Self-Organization},
  journal = {Chaos: An Interdisciplinary Journal of Nonlinear Science},
  volume  = {17},
  number  = {2},
  pages   = {026103},
  year    = {2007},
  doi     = {10.1063/1.2737822}
}

@article{hemmati2013,
  author  = {R. Hemmati and R.-A. Hooshmand and A. Khodabakhshian},
  title   = {State-of-the-Art of Transmission Expansion Planning: Comprehensive Review},
  journal = {Renewable and Sustainable Energy Reviews},
  volume  = {23},
  pages   = {312--319},
  year    = {2013},
  doi     = {10.1016/j.rser.2013.03.015}
}

@article{tinney1967,
  author  = {W. F. Tinney and C. E. Hart},
  title   = {Power Flow Solution by {N}ewton's Method},
  journal = {IEEE Transactions on Power Apparatus and Systems},
  volume  = {PAS-86},
  number  = {11},
  pages   = {1449--1460},
  year    = {1967},
  month   = {nov},
  doi     = {10.1109/TPAS.1967.291823}
}

@article{matpower2011,
  author  = {R. D. Zimmerman and C. E. Murillo-S\'{a}nchez and R. J. Thomas},
  title   = {{MATPOWER}: Steady-State Operations, Planning, and Analysis Tools for Power Systems Research and Education},
  journal = {IEEE Transactions on Power Systems},
  volume  = {26},
  number  = {1},
  pages   = {12--19},
  year    = {2011},
  month   = {feb},
  doi     = {10.1109/TPWRS.2010.2051168}
}
